\documentclass[11pt,a4paper,english]{article}
\usepackage{jheppub}

\usepackage{xcolor}
\usepackage{changebar}

\colorlet{ins}{blue}
\colorlet{del}{red}

\usepackage[backend=bibtex8,url=false,eprint=true,doi=true,sorting=none,backref=true]{biblatex}
\usepackage{ifpdf}
\usepackage[T1]{fontenc}

\usepackage{amsmath,amssymb,amsfonts,amsthm}

\usepackage{epsfig}
\usepackage{graphicx}
\usepackage{latexsym}
\usepackage{color}
\usepackage[colorinlistoftodos, shadow, disable]{todonotes}
\usepackage{caption}
\usepackage{subfig}

\makeatletter

\@ifundefined{textcolor}{}
{%
 \definecolor{BLACK}{gray}{0}
 \definecolor{WHITE}{gray}{1}
 \definecolor{RED}{rgb}{1,0,0}
 \definecolor{GREEN}{rgb}{0,1,0}
 \definecolor{BLUE}{rgb}{0,0,1}
 \definecolor{CYAN}{cmyk}{1,0,0,0}
 \definecolor{MAGENTA}{cmyk}{0,1,0,0}
 \definecolor{YELLOW}{cmyk}{0,0,1,0}
 \definecolor{PURPLE}{rgb}{0.5,0,0.5}
 }

\newcommand{\bite}{\begin{itemize}}
\newcommand{\eat}{\end{itemize}}
\newcommand{\beq}{\begin{equation}}
\newcommand{\eeq}{\end{equation}}

\newcommand{\beqa}{\begin{align}}
\newcommand{\eeqa}{\end{align}}
\newcommand{\barr}{\begin{array}}
\newcommand{\earr}{\end{array}}

\newcommand{\mb}[1]{\mathbf{#1}}
\newcommand{\mc}[1]{\mathcal{#1}}

\newcommand{\mf}[1]{\mathfrak{#1}}

\newcommand{\vect}[1]{\boldsymbol{\mathrm{#1}}}

\newcommand{\fullket}[1]{\vert #1 \rangle}

\newcommand{\sltwoc}{SL(2,\mathbb{C})}
\newcommand{\sltwoz}{SL(2,\mathbb{Z})}

\newcommand{\utilde}[1]{\underset{\sim}{#1}}

\newcommand{\ignore}[2]{\hspace{0in}#2}

\makeatother

\title{Quantum Hall Effect and Black Hole Entropy in Loop Quantum Gravity}

\author{Deepak Vaid}
\emailAdd{dvaid79@gmail.com}
\preprint{}

\abstract{In LQG, black hole horizons are described by $2+1$ dimensional boundaries of a bulk $3+1$ dimensional spacetime. The horizon is endowed with area by lines of gravitational flux which pierce the surface. As is well known, counting of the possible states associated with a given set of punctures allows us to recover the famous Bekenstein-Hawking area law according to which the entropy of a black hole is proportional to the area of the associated horizon $ S_{BH} \propto A_{Hor} $. It is also known that the dynamics of the horizon degrees of freedom is described by the Chern-Simons action of a $\mathfrak{su(2)}$ (or $\mathfrak{u(1)}$ after a certain gauge fixing) valued gauge field $A_{\mu}^i$. Recent numerical work which performs the state-counting for punctures, from first-principles, reveals a step-like structure in the entropy-area relation. We argue that both the presence of the Chern-Simons action and the step-like structure in the entropy-area curve are indicative of the fact that the effective 
theory which describes the dynamics of punctures on the horizon is that of the Quantum Hall Effect.}

\begin{document}

\date{\today}

\maketitle

\listoftodos

\tableofcontents

\begin{quotation}
\emph{And I cherish more than anything else the Analogies, my most trustworthy masters.
They know all the secrets of Nature, and they ought least to be neglected in Geometry.}
\flushright{- Johannes Kepler}
\end{quotation}

\ignore{
\section{Reference Outline}\label{sec:outline}

\begin{enumerate}
	\item \textbf{Area Operators}: Spin-network edges which don't end of a vertex correspond to fermions (Rovelli + Morales). Schematic picture is that of open strings floating around, with unattached ends representing fermions which endow an abstract triangle with a unit of area.
	\item We want to consider how these fermions can pair up. Picture is that of four edges meeting to form a vertex at a tetrahedron/sphere.
	\item Presence of fermions in theory also leads to non-zero value for $ e \wedge e $ term via integration of Gauss' constraint on a 2-sphere when we start with an action for gravity with fermionic matter. Thus fermions source the area.
	\item Why four edges? Two reasons a). Penrose-Teller says that a uniform sphere appears uniform to boosted observers. Therefore it must have defects. Three points can be mapped to a single great circle (stereographic map -> conformal transformation). Four points with non-zero cross-ratio cannot!
        \item Second reason b). Four is the minimum number of faces needed for a non-trivial quantum of 3-volume, as shown by ... [Renate Loll?]
        \item Hall effect requires that the translational invariance of the surface be broken. Otherwise the Hall effect will not occur - one can always boost to a frame where the force on the charge carriers due to the applied magnetic field is zero. Translational symmetry of the surface is broken due to presence of punctures as shown in \cite{Kabat1994Canonical}.
	\item We can consider tetrahedron or sphere. It doesn't matter. What matters is the behaviour of the 2-sphere under symmetries exchanging pairs of punctures.
        \item Boundaries can be sown together by mapping microstate on one surface to microstate on another.
        \item Diffeomorphism invariance of bulk is required, w.r.t ``small'' diffeos (usual co-ordinate transformations) \emph{and} w.r.t ``large'' diffeos generated by discrete symmetries of the surface states.
        \item States invariant under discrete symmetries are constructed using solutions of the Yang-Baxter equation.
        \item Horizon microstates are identified with quantum hall states (content of qhe\_bhe project)
        \item \emph{Physical} states space of quantum 3-geometry consists of surface wavefunctions of the quantum hall form, glued together with preonic excitations!
\end{enumerate}
}

\section{Introduction}\label{sec:intro}

The problem of black hole entropy (BHE) is one of the most outstanding and vexing questions in modern physics. It lies at the interface of the two pillars of contemporary theoretical physics: General Relativity (GR) and Quantum Field Theory (QFT). The Quantum Hall Effect (QHE) lies at the frontier of condensed matter research with possibilities of exciting future applications in fields such as Quantum Computing. It is a many-body phenomena which is observed in measurements of the Hall voltage across a strip of material subjected to extremely low temperatures and very high magnetic field strengths (\autoref{fig:qhe_setup})\footnote{The Hall conductivity $\sigma_{xy}$ corresponds to the off-diagonal component of the conductivity tensor}. As such the QHE is diametrically opposite to the question of BHE in the sense that it is now routinely observed in laboratories around the world, whereas we are yet to encounter black holes in anything other than our theoretical imaginations or in far reaches of the cosmos at 
the centers of galaxies where the observational evidence is again indirect. If however, there was to exist a physical analogy between these widely separated phenomena then the possibility arises that conjectures about black hole physics could be tested in an experimental setting. In the this paper we will argue for the existence of a compelling correspondence between BHE and the QHE which could help in bringing this objective closer to fruition.

The layout of this paper is as follows. In the rest of this section we sketch an outline of the holographic principle (\autoref{subsec:holography}) and motivate the assertion that quantum geometry is best understood by analogy with many-body phenomena in condensed matter physics (\autoref{subsec:emergent}). In \autoref{sec:hall_effect} we describe the essentials of the quantum hall effect. In \autoref{sec:state-counting} we describe the step-like structure of the black hole entropy revealed by \cite{Corichi2006Quantum,Corichi2007Black}. In \autoref{subsec:spin-nets} we summarize the construction of the area operator in LQG in terms of punctures on a horizon and see how we can exploit this fact to perform an exact counting of the micro-states of quantum geometry of an isolated horizon. In \autoref{sec:linking} and \autoref{sec:hubbard} we establish the correspondence between the physics of the hall effect and that of the horizon degrees of freedom and use this fact to write down a candidate non-perturbative 
many-body ground 
state for quantum gravity. Finally in \autoref{sec:conclusion} we conclude with a summary of our results and outline directions for future research.

\subsection{Holography}\label{subsec:holography}

There is by now a vast amount of literature on the AdS/CFT correspondence. The ``AdS'' part refers to the physics of a $ d+1 $ dimensional spacetime $\mathcal{M}$ whose geometry approaches that of anti-deSitter spacetime \emph{asymptotically}, i.e. as one approachs the boundary at null infinity. AdS is a \emph{compact} manifold with boundary. Therefore there is a notion of points being near the boundary or far from the boundary and deep in the bulk. The requirement that the spacetime metric approachs the AdS form only asymptotically - i.e. for observers close to the boundary - leaves us with a lot of freedom to specify the interior geometry which can be distinct from the AdS bulk metric. In particular one can have various sorts of black holes - such as Schwarzschild, black branes, etc. - embedded in the interior of $\mc{M}$. The ingredients of the first version of the correspondence \cite{Maldacena1998The-Large} consisted of a five-dimensional Anti-deSitter spacetime along with five microscopic dimensions curled up or \emph{compactified} into the topology of a 
five-sphere $ S_5 $. The complete spacetime is then written as: $AdS_5 \times S_5$.

This spacetime arises as the low-energy limit of a certain superstring theory (Type IIB) which turns out to be a supergravity theory (``SUGRA'') living in ten dimensions accompanied by various scalars and gauge fields coming from the non-perturbative sectors of string theory (NS-NS,Ramond-Ramond etc.). The number of macroscopic dimensions can be reduced to five by compactifying five of the dimensions into a manifold ($ S_5 $) whose microscopic extent is determined by the string scale. Consequently the boundary of the $AdS_5 \times S_5$ spacetime also consists of four macroscopic dimensions corresponding to the five which describe the bulk.

The symmetry group of the boundary of an anti-deSitter spacetime is the conformal group \cite{Wald1984General}. Thus any field theory defined on the boundary of an AdS spacetime must be a conformal field theory. In the $AdS_5 \times SO(5)$ case, the boundary CFT is a $\mc{N}=4$ super(symmetric) Yang-Mills theory\footnote{$\mc{N}$ is the number of supersymmetry charges. $\mc{N}=4$ is the maximum for super Yang-Mills in four dimensions.}. The AdS/CFT correspondence then provides us with a dictionary for computing correlation functions for matter field in the bulk spacetime by equating them with expectation values of the asymptotic values of the relevant field operators on the CFT boundary \cite{Maldacena1998The-Large,Gubser1998Gauge,Witten1998Anti}.

Later work has shown that the correspondence is far more general and the requirement of an asymptotically anti-deSitter bulk can be loosened to include deSitter and even FRW geometries and, in principle, the dimensionality of the bulk $d$ can be arbitrary. Thus we will henceforth refer to this duality simply as \emph{holography} since it relates physics in a $d+1$ dimensional spacetime with the physics of its $d$ dimensional boundary, in a manner reminiscent of the way in which a three-dimensional image can be encoded into two-dimensional ``holograms''.

\subsection{Emergent Geometry}\label{subsec:emergent}

\ignore{The dream of constructing a physically viable model of macroscopic geometry as an emergent phenomenon has long been pursued. }Simple models of dynamical systems in cosmology such as those built around the LFRW and other symmetry-reduced spacetimes are inadequate for a description of the universe as we observe it today. From a quantum geometry perspective a spacetime is best described as consisting of elementary quanta of geometry and matter glued together to form a structure which, on large enough scales, resembles a smooth geometry. Gravity is then best thought of as a many-body phenomena. At the very least that is the author's prejudice.

LeMaitre-Friedmann-Robertson-Walker (LFRW) and similar models attempt to make Einstein's equations tractable by reducing the number of geometric degrees of freedom from an uncountable infinity to a handful by imposing restrictions of homogeneity and isotropy on the metric. Such \emph{single-body} models can go only so far as to provide a hint of things to come as one cannot reconcile them with the multitude of astronomical observations which point towards an accelerating universe (i.e. with non-zero $\Lambda_{eff}$) which is extremely homogeneous and isotropic while at the same time allowing for the formation of complex hierarchy of structures within this uniform landscape.

LFRW-like models which, with the addition of scalar and other fields, allow us to compute the spectrum of density perturbations produced during an inflationary period, are an inherently phenomenological construct in that they do not shed light on the microscopic \emph{many-body} dynamics that is the correct description of the background geometry. The scalar field usually employed in these models represents only the effective interaction of matter with geometry. The background geometry in these models is characterized by the time-dependence of the scale factor \footnote{In the simplest case the metric contains only one free parameter - the scale factor $a$. One can construct anisotropic metrics however the same limitation applies to these too}. From the perspective of quantum geometry the background geometry in such models is only an ``effective'' or mean-field solution of the underlying discrete, many-body system.

What is hidden under the rug, so to speak, is that the background geometry is a system with its own degrees of freedom. The evidence for this comes from two observations. First from the Hamiltonian formulation of general relativity \cite{Thiemann2001Introduction} we know that each point in 3+1 dimensional space-time the metric has exactly two degrees of freedom for the case of gravity without matter\footnote{The interested reader can refer to the argument on pg. 266 of \cite{Wald1984General}}. Secondly, From developments in String Theory, Quantum geometry and the study of the question of black hole entropy we have realized that a given volume of spacetime $\Sigma$ with boundary $\partial\Sigma$ can contain only a finite amount of information proportional to the area of $\partial\Sigma$. Taking this information to be represented by a spinor in each unit-cell of a microscopic geometrical state satisfies both these facts. While this physical picture of a discrete geometry emerges from both LQG and String Theory 
(in some of its incarnations), what has so far been missing is a satisfactory mechanism for reconstructing an effective classical geometry from this bare skeletal structure.

In this work we argue that such systems have analogs in condensed matter many-body physics whose properties have already been extensively studied over the past few decades. The insights gained from the study of the integer and fractional quantum hall effects can then be directly to the question of black hole entropy.

Let us note that a great deal of work along similar lines has been done previously. Various authors \cite{Hartnoll2008Building,Hartnoll2008Holographic,Hartnoll2010Lectures,Fujita2012SL2Z,Bayntun2011Finite,Jokela2012Fluctuations,Jokela2011A-holographic,Bergman2010Quantum,Bayntun2010AdS/QHE:,Albash2010Landau,Albash2008A-Holographic} have investigated holographic descriptions of the QHE. The closest to our current approach is the work done by Jokela et al. who work with string theory inspired backgrounds containing D2 and D8 branes. Our model is simpler conceptually in that it does not need to invoke string theory and nor are we concerned with extra dimensions. 

generally written as a three-dimensional vector $ \vect{L} $:
%
%



\section{Quantum Hall Effect}\label{sec:hall_effect}

The discovery of the integer and fractional quantum hall effects (IQHE/FQHE) is one of the few truly exciting developments in physics in the past three decades. The QHE is the umbrella moniker given to a variety of quantum many-body systems consisting of two-dimensional electron gases which when exposed to strong external magnetic fields $\vect{B}$ perpendicular to the plane containing the system, exhibit incredibly precise quantization of transport coefficients. Such two-dimensional electron gases (2DEG) can be realized in a variety of ways, of which some of the most frequently used methods are based on interfaces between GaAs and GaAl semiconductor heterostructures.

\begin{figure}[htbp]
	\begin{center}
	\subfloat{
      		\includegraphics[scale=0.30]{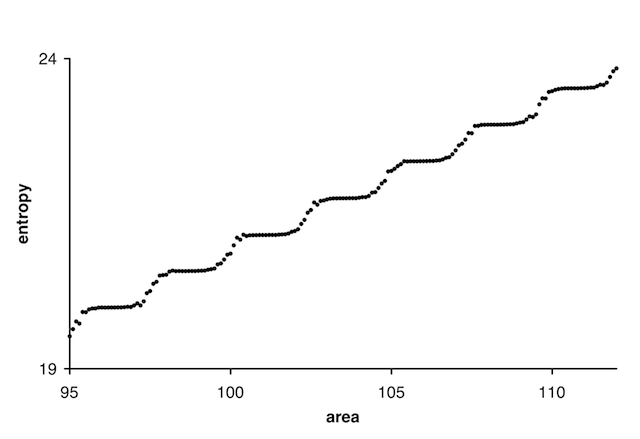}
		\label{fig:bhe_corichi}
	}
	\subfloat{
		\includegraphics[scale=0.3]{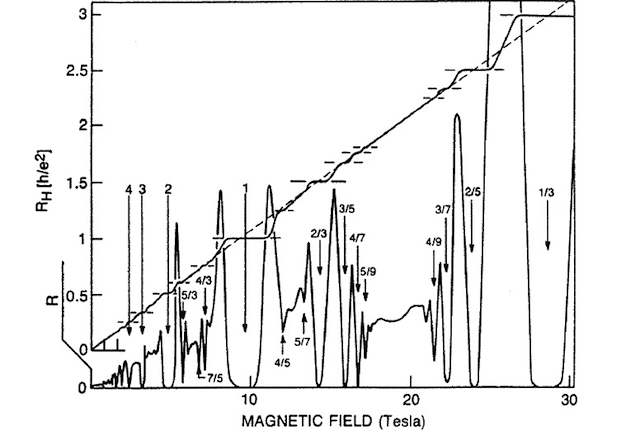}
      		\label{fig:fqhe_data}
	}
	\caption{Comparing black hole entropy to the quantum hall effect. An exact calculation of the entropy of an isolated horizon in lqg (\emph{left}) showing the step-like dependence of the entropy on the area of the horizon. An experimental trace of the longitudinal (or \emph{dissipative}) $\rho_{xx}$ and transverse (or \emph{hall}) $\rho_{xy}$ resistivity (\emph{right}, source: \cite{Stormer1992Two-dimensional}). As the magnetic field is varied, while keeping the electron density fixed, the number of magnetic flux quanta through the sample changes, causing the filling fraction $\nu$ to change. The Hall resistance is precisely quantized at ``plateaus'' for certain integer and fractional values of $\nu$, while $\rho_{xx}$ vanishes in those regions indicating the presence of superflow in the longitudinal direction.}
    \label{fig:bhe_qhe_comparison}
    \end{center}
\end{figure}

For certain intervals of the magnetic field a dramatic shift is observed in the behavior of transport coefficients such as the longitudinal or \emph{dissipative} resistivity $\rho_{xx}$ (see e.g., \cite{MacDonald1994Introduction,Goerbig2009Quantum}) and the transverse or \emph{Hall} resistivity $\rho_{xy}$. As the magnetic field is increased, for certain intervals, the Hall resistance $\rho_{xy}$ is \emph{exactly} quantized in units of $h/e^2$, whereas the dissipative resistance goes to zero (\autoref{fig:bhe_qhe_comparison}, right-hand side) - indicating the formation of superflow in the longitudinal current. The remarkable aspect of this quantization is its robustness with respect to the precise microscopic form of the interparticle interactions, the size and shape and the amount of disorder in the experimental samples.

\subsection{Classical Hall Effect}\label{subsec:classical-hall-effect}

\begin{figure}[htbp]
	\centering
	\includegraphics[scale=0.6]{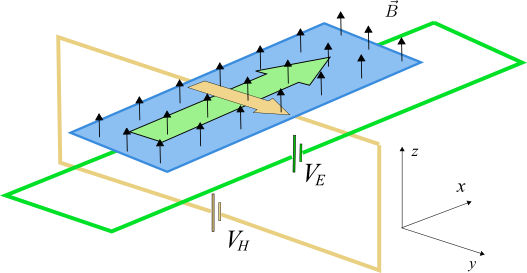}
	\caption{Schematic of the experimental setup for the quantum hall effect. The black arrows represent the magnetic field in the z-direction, perpendicular to the sample. The applied e.m.f is along the x-axis (green arrow) and the resulting Hall current (yellow arrow) is in the y direction.}
	\label{fig:qhe_setup}
\end{figure}

Let us review the classical Hall effect. It occurs as a consequence of the Lorentz force law which governs the behavior of charged particles in the presence of external magnetic fields:
\begin{equation}\label{eqn:lorentz_force}
	\vect{F} = \vect{E} + q \, \vect{v} \times \vect{B}
\end{equation}
The simplest manifestation of this law is in the Hall effect in which a two-dimensional conductor through which a current is flowing due to an external e.m.f. $V_E$ is exposed to a transverse magnetic field as depicted in \autoref{fig:qhe_setup}. The magnetic field causes the electron current $\vect{j}$ to develop a non-zero transverse component $j_x \ne 0$ leading to the formation of a potential difference - the ``hall potential'' $V_H$ between the two-sides of the conductor.

The Hall resistance $R_H = \rho_{xy}$ is given by the ratio of the Hall potential and the longitudinal current:
\begin{equation}
	R_H = \frac{V_H}{I}
\end{equation}
According to the Drude theory of magnetoelectric transport, for the case of a DC current, the equation of motion of charge carriers with momentum $\vect{p}$ is given by \cite{Goerbig2009Quantum}:
\begin{equation}\label{eqn:drude}
	\frac{d\vect{p}}{dt} = -e \left( \vect{E} + \frac{\vect{p}}{m_b} \times \vect{B}\right) - \frac{\vect{p}}{\tau}
\end{equation}
where $\vect{E},\vect{B},-e,\tau, m_b$ are the electric and magnetic fields, the electron charge, the band mass and the relaxation time for scattering processes, respectively. At equilibrium we have $d\vect{p}/dt = 0$. For 2d electrons with $\vect{B} \equiv B_z$ the e.o.m becomes:
\begin{subequations}\label{eqn:drude2}
	\begin{align}
		e E_x & = -\omega_c p_y - \frac{p_x}{\tau} \\
		e E_y & = \omega_c p_x - \frac{p_y}{\tau}
	\end{align}
\end{subequations}
where $\omega_c = \frac{e B_z }{m_b}$ is the cyclotron frequency which determines the radius of the orbit of a charged particle in a perpendicular magnetic field.

Now, in linear response theory the resistivity tensor $\utilde{\rho} \equiv \rho_{ab}$ is defined:
\begin{equation}\label{eqn:resistivity-def}
	\vect{E} = \rho \vect{j}
\end{equation}
Expressing the current density in terms of the momentum:
\begin{equation}
	\vect{j} = - \frac{e n}{m_b} \vect{p} 
\end{equation}
where $n$ is the local electron density in the sample. Thus \eqref{eqn:drude2} becomes:
\begin{subequations}\label{eqn:drude3}
	\begin{align}
		E_x & = \frac{1}{\sigma_0} \left( \omega_c \tau j_y + j_x \right)  \\
		E_y & = -\frac{1}{\sigma_0} \left( \omega_c \tau j_x + j_y \right)
	\end{align}
\end{subequations}
where $\sigma_0 = ne^2\tau/m_b$ is the Drude conductivity. From the above we can read out the conductivity tensor:
\begin{equation}\label{eqn:resistivity-tensor}
	\utilde{\rho} = \frac{1}{\sigma_0}\left( \begin{array}{cc} 1 & \omega_c \tau \\ -\omega_c \tau & 1 \end{array} \right) = \frac{1}{\sigma_0}\left( \begin{array}{cc} 1 & \mu B \\ -\mu B & 1 \end{array} \right)
\end{equation}
where $\mu = e \tau/m_b$ is known as the \emph{mobility}. The hall resistivity is the off-diagonal component $\rho_{xy}$ of the resistivity tensor:
\begin{equation}\label{eqn:hall-resistance}
	R_H = \rho_H = \frac{\omega_c \tau}{\sigma_0} = \frac{B}{n e c}
\end{equation}
where $e$ is the electron charge, $n$ is the density of electrons and $c$ is the speed of light. For a sample in which $n$ is constant the Hall resistance is a linear function of the external magnetic field.

The conductivity tensor $\utilde{\sigma} $ is obtained by taking the inverse of the resistivity tensor:
\begin{equation}\label{eqn:conductivity-tensor}
	\utilde{\sigma} = \utilde{\rho}^{-1} = \frac{\sigma_0}{1 + \mu^2 B^2} \left( \begin{array}{cc} 1 & -\mu B \\ \mu B & 1 \end{array} \right) = \left( \begin{array}{cc} \sigma_L & -\sigma_H \\ \sigma_H & \sigma_L \end{array} \right)
\end{equation}
where $\sigma_L = \sigma_0/(1+\mu^2 B^2)$ is the longitudinal conductivity and $\sigma_H = \mu B \sigma_0/(1+\mu^2 B^2)$ is the Hall conductivity. In the limit of no scattering ($\tau \rightarrow \infty, \mu B \rightarrow \infty$), $\sigma_L \rightarrow 0$ and the conductivity tensor becomes:
\begin{equation}\label{eqn:conductivity-tensor2}
	\utilde{\sigma} = \left( \begin{array}{cc} 0 & -\sigma_H \\ \sigma_H & 0 \end{array} \right)
\end{equation}

\subsection{Landau Levels}\label{subsec:landaulevels} 

The hall effect hinges on our ability to understand the behavior of electrons in two-dimensions in the presence of transverse (orthogonal to the plane containing the electron fluid) magnetic fields. Equation \ref{eqn:lorentz_force} tells us that in the absence of an external electric field, an electron, of mass $m_e$, moving at velocity $\vect{v}$ experiences a centrifugal force:
\begin{equation}\label{eqn:centrifugal}
	F_r = e v B = \frac{m_e v^2}{r}; \qquad (v = |\vect{v}|)
\end{equation}
causing it to move in circular orbits with a radius given by
\begin{equation}\label{eqn:cyclotron-radius}
	r_c = \frac{m_e v}{e B}
\end{equation}
Classically the momentum, and hence the radius, of an electron in a circular orbit can be any positive real number. However in strong magnetic fields, where a quantum mechanical analysis is necessary, electron orbits are quantized much like the energy level of a harmonic oscillator. Electrons are constrained to move in orbits with energy:
\begin{equation}
	E_n = \hbar \omega_C \left( n + \frac{1}{2} \right)
\end{equation}
where $n \in \mb{Z}$ is a quantum number labeling the allowed levels.

\begin{figure}[htbp]
\begin{center}
\includegraphics[scale=0.4]{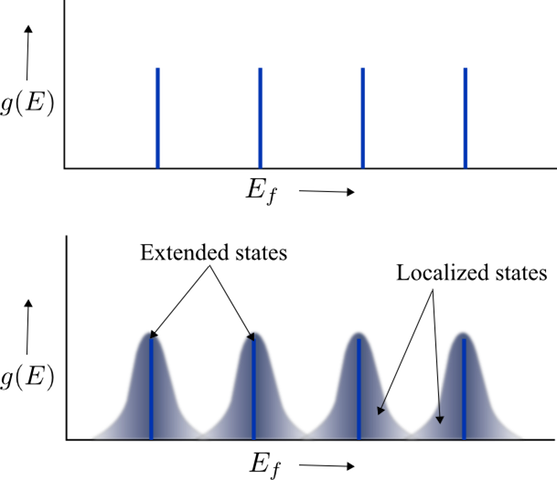}
\end{center}
\caption{Landau levels of a 2DEG in a clean (top) and in a disordered (bottom) sample. In the absence of disorder Landau levels occur at sharply defined intervals of the cyclotron energy $\hbar\omega_c$. Disorder due to defects and impurities leads to localization of electron wavefunctions in the sample and the broadening of Landau levels.}
\label{fig:landau_levels}
\end{figure}

\begin{figure}[htbp]
\centering
\includegraphics[scale=0.4]{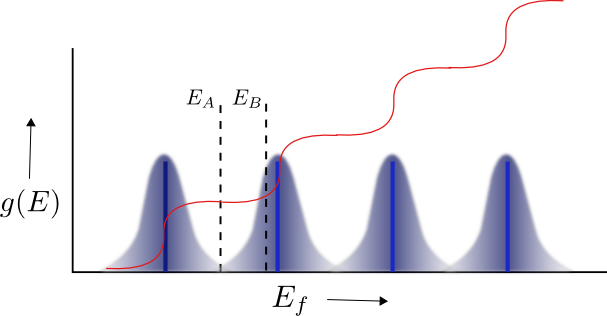}
\caption{The relation between the step-like change in conductivity as the Fermi energy $E_f$ is varied and the density of states at that energy is shown. $E_A$ and $E_B$ are two illustrative cases. The efficiency with which charge can be transported through the system depends on the presence and number density of \emph{extended} states which occur in the vicinity of the Landau levels. The states in the tails owe their existence to disorder and are \emph{localized} which, therefore, cannot contribute to transport coefficients.}
\label{fig:landau_levels_2}
\end{figure}

\subsection{Filling fraction}\label{subsec:filling-fraction}

At this point it is useful to define a quantity called the ``filling factor'' or $\nu$ which is the ratio of the number of electrons to the number of magnetic flux quanta in the sample:
\begin{equation}
	\nu = \frac{\rho}{B/\phi_0}
\end{equation}
where $\phi_0 = hc/e $ is the quantum of magnetic flux. In terms of $\nu$ the Hall resistance becomes:
\begin{equation}
	R_H = \frac{\phi_0}{\nu ec} = \frac{h}{\nu e^2}
\end{equation}

\subsection{Flux Attachment}\label{subsec:fluxattachment}


In the fractional quantum hall effect (FQHE) we observe plateaus at \emph{fractional} values of the filling factor $\nu$. Here, the above mentioned physical interpretation of $\nu$ becomes significant. If $ \nu$ is a fraction of the form $n/(2pn \pm 1)$, where $p$ and $n$ are integers, then according to the \emph{composite fermion} theory the corresponding FQHE state can be mapped to an effective IQHE interactions between composite fermions which consist of an electron paired up with an even number of flux quanta.

The addition of a Chern-Simons term to the effective action of quantum hall system has the effect of attaching a unit flux of the gauge field strength with each electron. Classically this has the effect of modifying the connection and hence the symplectic structure on the phase space. Quantum mechanically this is accomplished by multiplying the exact n-electron state by a - possibly non-abelian - phase factor.

It is important to stress that neither the classical nor the quantum hall effect requires a the presence of a \emph{net} flux $ \Phi $  through the sample surface. It is sufficient to have a \emph{locally} non-vanishing magnetic flux density. This can be accomplished by making the black hole magnetically charged. Both a monopole and a dipole configuration would lead to a non-zero local magnetic flux density. However, for a black hole with a magnetic monopole the net flux $ \Phi \ne 0 $ whereas for a magnetic dipole $ \Phi = 0 $.


\section{Black Hole Entropy}\label{sec:state-counting}

It is now well understood that isolated horizons describe a large class of black holes including those with electric, magnetic and possibly Yang-Mills charges and with or without rotation. Isolated horizons satisfy the following \emph{generalized second law} of thermodynamics:
\begin{equation}
\delta E = \frac{k}{8\pi} \delta A + \Omega \delta J + \Phi \delta Q
\end{equation}
where $k, \Omega, \Phi$ are the surface gravity, the angular velocity and the electrostatic potential, respectively. $ A, J, Q$ are the area, the angular momentum and the electric charge, respectively. The most general way to write the above expression is:
\begin{equation}
\delta E = \sum_i \vect{p}_i d \vect{q}_i
\end{equation}
where $\{\vect{p}_i, \vect{q_i} \}$ are canonical momenta and co-ordinates for all the various macroscopic dynamical observables associated with a general isolated horizon.

with $ J, Q = 0 $ for the uncharged case:
\begin{equation}
\delta E = \frac{k}{8\pi} \delta A
\end{equation}

\subsection{Step-like structure of black hole entropy}

As shown in the numerical work of Corichi, et al. \cite{Corichi2006Quantum,Corichi2007Black} the entropy S of a black hole as a function of its area A exhibits plateaus which are very reminiscent of the plateaus in the hall resistance RH as a function of the magnetic field B in the Quantum Hall Effect (QHE). For comparison we display both graphs in \autoref{fig:bhe_qhe_comparison}.

The graph on the left is that from Corichi et al's numerical computation of the black hole entropy ($S_{bh}$) by an exact counting of the microstates corresponding to the area eigenvalue $A$. The one of the right shows the response of the hall resistance ($R_H$) as a function of the transverse magnetic field ($B$). The similarity between the two graphs is more than skin deep. The simple fact that the entropy function has the ladder shape tells us a great deal. Let us denote the value of area where a jumps occurs as $a_i$ and $s_i$ as the (almost constant) entropy between two consecutive jumps $a_i$ and $a_j$. Knowing the Boltzmann equation relating the entropy to the logarithm of the number of states of a system, and from the shape of the entropy-area function, we can guess that the form of density of states functional of the horizon as a function of area will be of the form shown in Fig. 3 and Fig. 4.

Now we summarize the present understanding of area operators in quantum geometry which allows us to make this correspondence more concrete.

\subsection{Spin-Networks and Area Operators}\label{subsec:spin-nets}

Our starting point for constructing a correspondence between the BHE and QHE is the picture of black hole entropy as discovered via the LQG route. Here the microscopic states of quantum geometry are given by abstract graphs known as ``spin-network'' embedded in a pre-geometric manifold. On this graph $ \Gamma $ volumes and areas are defined to be expectation values of the corresponding operators. The edges of the graph are labeled by spins $j_{i}$ which live in the fundamental representation of $SU(2)$. Each vertex $ v \in \Gamma $ is labeled by an intertwiner. Consider a vertex $ v_i $ to which $ N $ edges are attached, $ k $ of which are ``ingoing'' and other $ N-k $ edges are ``outgoing''. Let us label the ingoing edges by spins $ \{ j_1 \dots j_k \}  $ and the outgoing ones by $ \{ j_{k+1} \dots j_N \}  $.  Then the intertwiner $I_v$ associated with the given vertex is map from the Hilbert space of the incoming spins to the Hilbert space of the outgoing spins at $ v $:
\begin{equation}
I_{a_1 \ldots a_k}{}^{a_{k+1}\ldots a_N}:\bigotimes_{m=1\ldots k}\mathcal{H}_{j_m} \rightarrow \bigotimes_{n=k+1\ldots N}\mathcal{H}_{j_n}
\end{equation}
Alternatively, an intertwiner can be thought of as an operator which acts on the edges coming into and going out from a vertex to yield a complex number $ z \in \mathbb{C} $:
\begin{equation}
I^{a_1 \ldots a_N}:\bigotimes_{i=1\dots N}\mathcal{H}_{i}\rightarrow\mathbb{C}\label{eqn:Intertwiner}
\end{equation}
An intertwiner serve to glue the edges at a vertex in a consistent manner. It serves to ensure that the polyhedron, enclosing a given vertex, whose faces are normal to the edges coming into the vertex and which have areas given by the Casimir of the spin $ j_{a_i} $ piercing that face, is closed.

Thus the intertwiner plays the role of a delta function $\delta \left(j_i  + \ldots + j_k - j_{k+1} + \ldots - j_{N} \right)$ which serves to ensure that momentum is conserved at any given vertex, in a manner analogous to analogous to the factor associated with each vertex in a momentum-space Feynman diagram $ \delta \left( p_1 + \dots + p_k - p_{k+1} - \dots - p_N \right) $. In the usual Feynman diagram picture we do not have any freedom to choose the form of this factor. Its existence is deemed necessary in order to ensure that momentum is neither created nor destroyed at any vertex in a Feynman diagram and its form is determined uniquely by the requirements of Lorentz invariance of the Minkowski spacetime. In a QFT calculation we implicitly  fix the Lorentz frame in which the amplitude for a given process is calculated. However in a successful theory of quantum geometry - one of whose hallmarks would be an emergent semiclassical geometry which obeys \emph{unbroken} diffeomorphism invariance - one cannot \
emph{a priori} fix such a frame for all the vertices in a graph. The independent degrees of freedom of the LQG intertwiner \cite{Engle2007Flipped, Engle2007The-loop-quantum-gravity} can be seen as encoding in the quantum theory the diffeomorphism invariance of classical General Relativity.

\begin{figure}[htbp]
\label{fig:surface_punctures}
  \begin{center}
    \includegraphics[width=0.48\textwidth]{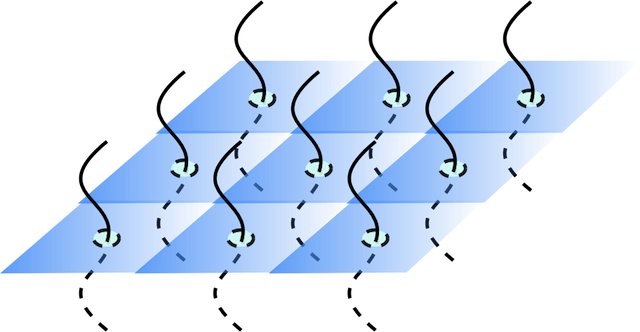}
  \end{center}
  \caption{An illustration of flux lines of the tetrad piercing a 2D surface thus endowing it with area at the locations of the punctures.}
\end{figure}


In this picture the surfaces are endowed by an area given by the casimir
of the spins of those edges which puncture the surface. Volumes are
given by expectation values of intertwiners operators corresponding
to the vertices contained within the prescribed region \cite{Rovelli1994Discreteness,Ashtekar1997cQuantum}.

\begin{figure}[htbp]
	\centering
	\includegraphics[scale=0.5]{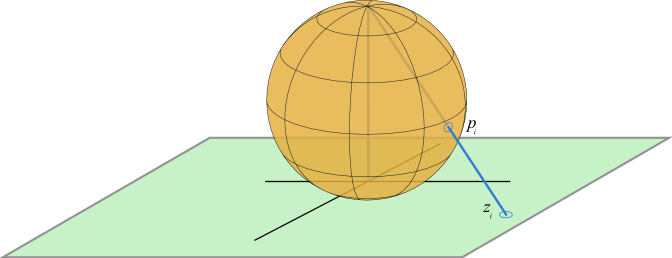}
	\caption{Stereographic projection of $S^{2}$ onto the complex plane $\mathbb{C}$}
	\label{fig:projection}
\end{figure}

The entire framework of LQG can be applied to the case when our manifold has internal boundaries in the form of isolated horizons \cite{Ashtekar1999Isolated}. In this case the graph is partitioned into two parts: one ($S_-$) identified with the ``interior'' of the horizon and the other ($S_+$) with the exterior. We do not have any information about what is within the sphere so we do not know what the graph looks like therein. The best we can do is characterize the surface itself in terms of the number of punctures of the exterior bulk spin-network $ S_+ $ which pierce it. The locations of the punctures on the surface can be given via stereographic projection onto the complex plane as shown in \autoref{fig:projection} and are labeled as $z_{1}\dots z_{4}$.

The projection onto $\mathbb{C}$ is shown in \autoref{fig:ComplexPlane}. The circle represents the boundary between the Northern ($ S^2_N $) and Southern hemispheres ($ S^2_S $) of the surface under the projection. What axis we choose to be our $ z $-axis for the projection is a gauge choice. We have chosen a projection such that $ z_1, \, z_2, \, z_3 $ and $ z_4 $ lie on $ S^2_N $ and $ S^2_S $ respectively.


\subsection{Cross Ratios and $\sltwoz$ Invariance}\label{subsec:punctures}

\begin{figure}[htbp]
	\centering
	\subfloat{
      	\includegraphics[scale=0.3]{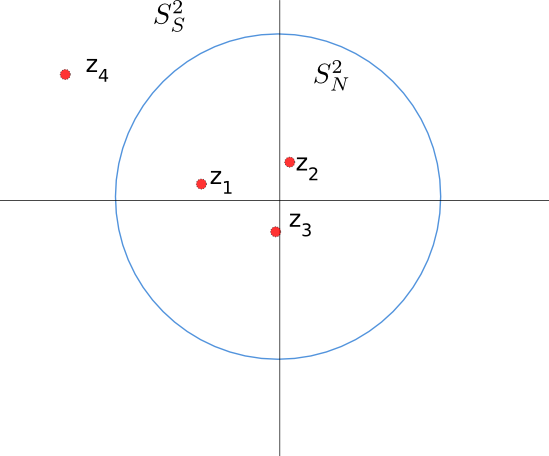}
		\label{fig:ComplexPlane}
	}
	\subfloat{
		\includegraphics[scale=0.3]{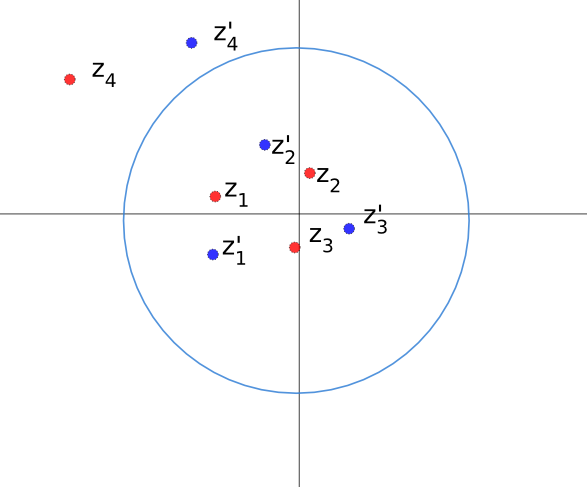}
		\label{fig:CoodTransform}
	}
	\caption{Mapping punctures to points on the complex plane $ \mathbb{C}^1 $. The circle denotes the boundary between the ``northern'' and the ``southern'' hemispheres, respectively. The effect of a lorentz transformation (rotation, boost or both) on the locations of the punctures.}
    \label{fig:projection_2}
\end{figure}

It was shown long ago by Penrose and Terrell, that a sphere does not appear to change its shape with respect to boosted observers. At first, this might be a counter-intuitive result. We are accustomed to the notion of length contraction of objects for moving obervers. Why then does the sphere not appear as an ellipsoid, with the major axis of rotation being transverse to the direction of motion? The reason for this is that for moving observers, apart from the length contraction, objects also appear to undergo what is referred to as a Penrose-Terrell \cite{Penrose1959The-apparent,Terrell1959Invisibility} rotation. For the case of a sphere, because of its symmetry, the length contraction is canceled by this rotation and it thus retains its shape for all moving observers.
The {}``inside'' and {}``outside'' regions of the sphere are described by the two portions of a spin-network as discussed above. For a semi-classical sphere we would require the surface to have a large number of punctures, with a certain distribution of spin-labels. Here we assume that the observer is suitably far away from the sphere so that its (the observer's) motion is in a flat background and is not subject to the gravitational field of the mass within the sphere. In order for the observer to characterize the geometry of the sphere we require the surface to have markings of some sort. The smallest number of markings is four, corresponding to the number of vertices of a tetrahedron which is the smallest simplex enclosing an element of 3-volume. These markings are labeled by the $z_{i}$ co-ordinates as mentioned earlier. Under a Lorentz transformation, these co-ordinates would appear to transform for an external observer to a new set: $z_{1}\dots z_{4}\rightarrow z'_{1}\dots z'_{4}$, which again take
values is $\mathbb{C}$ as shown in \autoref{fig:CoodTransform}.
This transformation can be represented by the action of an $\sltwoc$ matrix, as discussed in great detail in Chap 1. of \cite{Penrose1988Spinors}. \ignore{In order to make the paper self-contained the basic notation and methodology of this approach is discussed in Appendix \ref{app:modular-group}.} Before we go further let us introduce the notion of a cross-ratio.

The cross-ratio is a projective invariant well-known to geometers. For any four points it is given by:
\begin{equation}
\chi(z_{1},z_{2},z_{3},z_{4})=\frac{(z_{1}-z_{3})}{(z_{2}-z_{3})}\frac{(z_{1}-z_{4})}{(z_{2}-z_{4})}\label{eqn:CrossRatio}
\end{equation}
The significance of this quantity lies in the fact that under conformal transformations of $S^{2}$, any set of four points can be mapped onto another set only if both sets have the same cross-ratios. The general transformation which preserves $\chi$ is known as a M\"{o}bius transformation:
\begin{equation}
f(z)=\frac{az+b}{cz+d}\qquad where\quad a,b,c,d\in\mathbb{C}\quad and\quad ad-bc\neq0\label{eqn:MobiusTransform}
\end{equation}

The reason for a minimum of four punctures is now simple to explain. Ordinarily two field configurations on $S^{2}$ would be physically equivalent as long as they are related by an $\sltwoc$ transformation \footnote{This being the induced action on $S^{2}$ of Lorentz transformations of external oberservers \cite{Penrose1988Spinors} }. Now it is a well-known result of projective geometry that given two sets of three (or fewer) points on $S^{2}$ we can always find a conformal map which takes one set into the other. However a set of four points can be mapped into another set of four points, \emph{iff} both sets have the same cross-ratio. In other words we need a minimum of four punctures in order to have a non-degenerate co-ordinatization of the 2-sphere. The specification of four punctures on $S^{2}$ and their co-ordinates ${z_{1},z_{2},z_{3},z_{4}}$, under a given stereographic projection, breaks the continuous symmetry of the complex plane from the M\"{o}bius Group ($P\sltwoc$) to its 
discrete subgroup, the Modular Group ($P\sltwoz$), elements of which preserve the cross-ratio. Physically, this means that a state of the sphere with four punctures transforms for external observers according to $\sltwoz$ instead of $\sltwoc$, i.e. \textbf{Lorentz transformations are quantized}.

The breaking of Lorentz symmetry from $\sltwoc$ to $\sltwoz$ implies that the translational invariance of the two-dimensional surface is broken. This is necessary condition for the hall effect to occur\footnote{otherwise the observer can always boost to a frame where the relativistically induced electric field cancels out the Hall component $E_y$ of the electric field vector.}.


We have thus arrived at the conclusion that a consistent treatment of states of quantum geometry with respect to external observers requires that the symmetry group of spacetime be broken from $\sltwoc$ to $\sltwoz$. The argument was heuristic in nature and made no reference to the quantum hall effect or properties of strongly interacting systems. Remarkably, as it turns out, the quantum hall effect \emph{does indeed} exhibits an exact $\sltwoz$ symmetry (see for e.g., \cite{Fujita2012SL2Z,Burgess2000Particle-Vortex,Lutken1992Duality,Dolan1999Duality}). Given the transverse and longitudinal conductivities we can construct a single complex number containing both quantities:
\begin{equation}\label{eqn:complex-conductivity}
	\sigma = \sigma_{xy} + i \sigma_{xx}
\end{equation}
It turns out that the conductivities at various values of the Hall fractions are related to each other by $\sltwoz$ transformations\cite{Bayntun2010AdS/QHE:}:
\begin{equation}
	\sigma \rightarrow \sigma' = \frac{a \sigma + b}{c\sigma + d}
\end{equation}
where the $a,b,c,d$ are integers which satisfy $ad-bc=1$, with $c$ even.


\section{Linking the QHE and BHE}\label{sec:linking}

Before we proceed to establishing the analogy between the QHE and black hole physics let us note the similarity between two fundamental aspects of both phenomena. The physics of the Hall effect is determined by the requirement of gauge invariance. Similarly the dynamics of gravitational systems is determined by the requirement of diffeomorphism invariance which, when gravity is translated into the connection formulation, is recast into the gauge invariance of the Einstein-Hilbert action under $\sltwoc$ gauge transformations.

The four ingredients required for the understanding the dynamics of the quantum hall effect are as follows:
\begin{enumerate}
	\item \emph{charged particles on a two-dimensional surface}: the surfaces which we are interested in are furnished by the horizons of black holes which, in addition to mass and area, are charged under the external gauge field.
	\item \emph{non-zero gauge field}: which encodes the interaction of the charged particles with external electric and magnetic fields. Requires that the Einstein-Hilbert action be supplemented by the Maxwell term or its non-abelian counterpart, the Yang-Mills term. Here, since the Chern-Simons term is a function of the (restriction to the surface of the)gravitational connection $\vect{A}$, the bulk interaction term must also be a function of the full four-dimensional connection $A$.  Thus we \emph{do not} need to introduce any additional gauge fields or their interactions in the bulk!
	\item \emph{statistical gauge Chern-Simons field $A_\mu^i$}: which serves to bind flux quanta to charge carriers. In the first-order formalism, the Einstein-Hilbert action must be supplemented by a Chern-Simons term in the presence of boundaries in order to have a consistent symplectic structure \cite{Engle2009Black,Engle2010Black}. However, the presence of a Chern-Simons term implies that the dynamics of the ``puncture fluid'' on the surface is identical to the phenomenology of the quantum hall effect.
	\item \emph{disorder}: breaks the translational invariance of the two dimensional surface allowing the quantum hall effect to \emph{occur}. The Landau levels are highly degenerate consisting of a continuum of non-localized states. In the presence of disorder and impurities, localized states become available for electrons and the Landau levels broaden out into Landau ``bands'' (Fig. \ref{fig:landau_levels}) leading to the formation of the characteristic ``plateaus'' in hall conductivity measurements (Fig. \ref{fig:fqhe_data}). Here we do not need to introduce any additional sources of disorder as explained in Section \ref{subsec:punctures}.
\end{enumerate}

We begin with the action for general relativity in self-dual variables, which contains all three ingredients:
\begin{align}\label{eqn:full_action}
	S_{SD} = \int_{\mathcal{M}} d^4 x\, \epsilon^{\mu\nu\alpha\beta} \epsilon_{IJKL} \, e_{\mu}{}^I e_{\nu}{}^J F_{\alpha\beta}{}^{KL} - \Lambda \,\det(e) + \beta\,F_{\mu\nu}{}^{IJ} F^{\mu\nu}{}^{KL}\epsilon_{IJKL}
	 - \int_{\mathcal{\partial M}} \frac{k}{2\pi} Y_{CS}(\vect{A}) + j^\mu_i \vect{A}_\mu^i
	\ignore{\frac{1}{\varepsilon_0}g^{\alpha\mu}g^{\beta\nu}f_{\alpha\beta}f_{\mu\nu}}
\end{align}
\ignore{\frac{1}{2G} e^{AA'} \wedge e^B_{B'} \wedge F_{AB}[A] 
	& + \frac{\alpha}{2 \pi} F^{AB} \wedge F_{AB}}
where $A$ is the four dimensional gauge connection and $F$ is the corresponding field strength. $\vect{A}$ is the restriction of $A$ to the three dimensional boundary $\mc{\partial M}$. \ignore{, $a_\mu$ is the maxwell gauge field and $f_{\mu\nu} = \partial_\mu a_\nu - \partial_\nu a_\mu$ the corresponding field strength.}

The first term in \ref{eqn:full_action} is the usual Einstein-Hilbert action written in the connection formulation in terms of an $\sltwoc$ gauge connection and a frame field or \emph{vierbien} $e_\mu{}^I$. The second term is the cosmological constant term. The third term is the Yang-Mills term. The rest are the boundary terms of which $Y_{CS}$ is the Chern-Simons lagrangian given by:
\begin{equation}\label{eqn:chern-simons}
	Y_{CS}[\vect{A}] = \mathrm{Tr}\left[ \vect{A} \wedge \vect{d} \vect{A} + \frac{2}{3} \vect{A} \wedge \vect{A} \wedge \vect{A} \right]
\end{equation}
and the last term generates interactions between surface currents $j$ and the gauge field $\vect{A}$.
\ignore{Under large gauge transformations the CS term transforms as $\int Y_{CS} \rightarrow \int Y_{CS} + 8 \pi^2 n$, where $n$ is an integer.}
\ignore{
\subsection{Isolated horizon boundary conditions and the hall effect}
\label{subsec:boundary_conditions}

The fundamental starting point for the correspondence between the question of black hole entropy and the quantum hall effect is the following relationship \cite{Smolin1995Linking} between the gauge curvature $F^{AB}$ and the area operator $e^{AA'} \wedge e^B_{A'}$ on the $S^2$ inner boundaries of a manifold $M$:
\begin{equation}
\frac{1}{G} e^{A} \wedge e^B = \frac{(k-\alpha)}{2\pi} F^{AB}
\end{equation}
This constraint arises from the self-dual action for gravity as given by Jacobson and Smolin (1987,88)

The identification of the area operator with the magnetic field strength on the boundary lies at the heart of the physical analogy between the physics of the quantum hall effect and the black hole entropy problem.

Now the Chern-Simons term on the 2+1D boundaries arises due to the presence of a $ F \wedge F $ term (CDJ, or $ RR $ term \cite{Smolin1995Linking,Capovilla1991Pure}).  This can be seen by considering the variation of the $ F \wedge F $ term w.r.t the connection:
\begin{eqnarray}
\delta ( F^i \wedge F^j ) & = & \delta(F^i) \wedge F^j  - F^i \wedge \delta (F^j) \\
			       & = & D_{[A]} (\delta A^i) \wedge F^j  - F^i \wedge D_{[A]} (\delta A^j) \\
			& = & 2 D_{[A]} (\delta A^i \wedge F^j ) - 2 \delta A^i  \wedge D_{[A]} ( F^j)
\end{eqnarray}
In the second line we have used the fact that $ \delta (F[A]^i) = F[A +\delta A]^i - F[A]^i = D_{[A]} (\delta A^i) $ and then apply the Leibniz rule for derivatives to obtain the third line. Integrating both sides of this expression over a 4-manifold then yields:
\begin{equation}
\delta \int d^4 x\, Tr( F^i \wedge F^j ) =  2 \int d^4 x\, D_{[A]} (\delta A^i \wedge F^j ) - 2  \int d^4 x\,  \delta A^i  \wedge D_{[A]} ( F^j)
\end{equation}
The second term on the right hand side contributes the $ D_{[A]} ( F^j) $ term to the e.o.m for the connection. Using Stokes' theorem the first term on the r.h.s. becomes:
\begin{equation}
\int_{M} d^4 x\, D_{[A]} (\delta A^i \wedge F^j ) = \int_{\partial M} d^3 x \,  (\delta A^i \wedge F^j ) \cdot \mathbf{n}
\end{equation}
This contribution is what makes the variational principle ill-defined. In order to fix this we introduce a boundary term into the action from the beginning, such that its variation will cancel this contribution. With the benefit of hindsight we know that the variation of a Chern-Simons term will satisfy this requirement:
\begin{equation}
S_{CS} = - \int_{\partial M} d^3 x \, Tr \left[ A \wedge dA + \frac{2}{3} A \wedge A \wedge A \right]
\end{equation}
whose variation gives:
\begin{equation}
\delta S_{CS} =  - \int_{\partial M} d^3 x \, Tr \left[ \delta A^i \wedge F^j \right]                                                                                     
\end{equation}
as can be checked by direct evaluation. Thus ``an action containing the $ F \wedge F $ term must be supplemented by the action of a Chern-Simons field on the boundary in order to yield a well-defined variational principle!'' This field is nothing more than the pullback of the full $3+1$ connection to the $2+1$ boundary.
}
\subsection{Non-zero cosmological constant $\Lambda > 0 $ as an external magnetic field}
\label{sub:cosmological-constant}

Now, as described in \autoref{sec:hall_effect}, the quantum hall effect has \emph{two} distinct gauge fields $a_{\mu}^i$ and $A_{\mu}^i$. $ a_\mu^i $ is the gauge field corresponding to an external magnetic field $\vec{B}$ penetrating the sample. $ A_\mu^i $ is the Chern-Simons or ``statistical'' gauge field which couples quanta of magnetic flux to the particle degrees of freedom, which are electrons in a 2DEG and punctures or \emph{conical singularities} in the case of a black hole horizon.

The (abelian) Chern-Simons Lagrangian density for the \emph{statistical} gauge field $A_\mu$, which is responsible for flux attachment to electrons and hence the formation of anyonic quasiparticles, is given by (\cite[pg. 11, Equations 3.17, 3.18]{Zhang1992The-Chern-Simons-Landau-Ginzburg}):
\begin{equation}\label{eqn:csterm-1}
 L_{CS} = \frac{e\pi}{2 \theta \phi_0} \epsilon^{\alpha\beta\gamma} A_\alpha \partial_\beta A_\gamma
\end{equation}
where $\theta/\pi = (2k+1)$ is the number of flux quanta attached to each electron, in terms of which we have:
\begin{equation}\label{eqn:csterm-2}
 L_{CS} = \frac{e}{(2k+1)2\phi_0} \epsilon^{\alpha\beta\gamma} A_\alpha \partial_\beta A_\gamma
\end{equation}
The (non-abelian) Chern-Simons Lagrangian density required for a consistent variational principle of the CDJ form of the gravitational action is of the form \cite[pg. 26, eqn. 50]{Smolin1995Linking}:
\begin{equation}\label{eqn:nonabelian-cs}
 L'_{CS}= - \frac{k'}{2\pi} \left[ A^i \wedge dA^i + \frac{1}{3} \epsilon_{ijk} A^i \wedge A^j \wedge A^k \right]
\end{equation}
where $k' = \frac{6\pi}{G^2 \Lambda} + \alpha$. $\alpha$ is the coefficient of the CP violating $F\wedge F$ term in the full action. In the event that we gauge fix the $\mf{su}(2)$ gauge field $A_\mu^i$ down to a $\mf{u}(1)$ gauge field $A^i$, the second term on the right-hand side of \eqref{eqn:nonabelian-cs} vanishes and we are left with an abelian Chern-Simons term:
\begin{equation}\label{eqn:abelian-cs}
 L'_{CS} = - \frac{k'}{2\pi} A^i \wedge d\,A^i 
\end{equation}
Comparing equations \eqref{eqn:abelian-cs} and \eqref{eqn:csterm-2} we see that we have a formal correspondence between the Chern-Simons coupling strength in the quantum hall effect and the strength of the Chern-Simons term in the gravitational case:
\begin{equation}
 \frac{6\pi}{G^2 \Lambda} + \alpha \equiv \frac{e}{(2k+1)2\phi_0}
\end{equation}
Setting $\alpha = 0$ for the time being, we see that the cosmological constant is proportional to the number of flux quanta attached to each electron:
\begin{equation}
 \Lambda \sim (2k+1)
\end{equation}
Increasing the strength of the external transverse magnetic field through the quantum hall sample increases $k$ and decreases the filling fraction $\nu = 1/(2k+1)$. This observation leads us to see that ``turning on'' the cosmological constant in the bulk can be understood as being equivalent to turning on the ``external'' magnetic field through the horizon. Flux quanta of this external field in turn bind with punctures on the horizon to form anyons. The step-like structure observed in the conductivity of the resulting fluid in then a manifestation of the IQHE.

\subsection{Quantized Conductivity}
\label{subsec:conductivity}
The hall effect involve charge transport on a two-dimensional surface in the presence of a transverse external magnetic field. The ``charges'' on a black hole horizon in the quantum geometry perspective are the punctures which correspond to quanta of area. In the general case we would consider a black hole with non-zero electric, magnetic and others charges \footnote{from the membrane paradigm \cite{Parikh1998An-Action,Damour2008String} we know that we are justified in thinking of all the charges in a black hole as being entirely on the horizon. If the interior of a black hole is to be identified as being in a superfluid or superconducting state then again thermal reasoning suggests that all the charge should be on the surface once the system reaches equilibrium.} In the simplest case we consider uncharged, non-rotating black holes - i.e. whose only thermal macroscopic observable is the horizon area $A$.

For such black holes the dynamics of punctures on the horizon is determined by coupling a current $j^i$ to the Chern-Simons \eqref{eqn:abelian-cs}:
\begin{equation}
 S_{cs} = - \frac{k'}{2\pi} \int_{\mc{\partial M}} A^i \wedge d\,A^i + A^i j_i
\end{equation}
Varying the above w.r.t the connection we find \cite[Sec 2.2]{Bayntun2010AdS/QHE:}:
\begin{equation}
	\frac{\delta S_{cs}}{\delta A^i} = 0 \Rightarrow j^i = \frac{k'}{2\pi} \epsilon^{ijk} F_{jk}
\end{equation}
Comparing with the definition of the conductivity tensor $ \vect{j} = \utilde{\sigma} \vect{E}$, where $\vect{E} := (E_x, E_y) = (F_{tx}, F_{ty})$, we can read of the conductivity tensor as:
\begin{equation}
	\utilde{\sigma} = \left( \begin{array}{cc} 0 & k'/2\pi \\ -k'/2\pi & 0 \end{array}\right)
\end{equation}
Since gauge invariance of the path integral restricts $k'$ to integral values\cite{Witten1989Quantum}, we find that the hall conductivity ($\sigma_{xy}$) is also quantized!

\section{A Local Hamiltonian for the Horizon}
\label{sec:hubbard}

A quantum state of horizon geometry is given by a collection of punctures on a two-dimensional surface. This information by itself is not sufficient to write down a Hamiltonian which describes the evolution of such surfaces. We might suspect that given the physical setting - \emph{a two-dimensional interacting collection of spins} - and the expected phenomenology - \emph{the relation between the black hole entropy to the quantum hall effect}, a natural choice for such system might be the Hubbard model or one of its closely related condensed matter kin.\todo{\small{two or three such ``kin'' - the XYZ, XX, ZZ etc. and the Kondo models, maybe?}}

\subsection{Lattice structure from holonomies}
\label{subsec:lattice-structure}

Lattices do not have a notion of distance in the most general setting and given an abstract set of punctures - on a surface which is moreover supposed to be \emph{null}(!), it is not clear how we can successfully map the punctures to spins on a lattice in a unique manner. To be able to do so we need, at the minimum, to construct a notion of distance between two punctures based purely on the topological information available to us via the non-zero holonomy in the form on a phase factor picked up by a puncture, located initially at the point $ z_i $ on the complex plane $ \mb{C} $, as it traverses a closed loop $ \Gamma $ which contains $ k $ other punctures [\textbf{illustration}] \todo{insert diagram of punctures traversing a loop ...} labeled by $ \{z_{i_1}, z_{i_2},\ldots,z_{i_k}\} $.

Schematically one would write:
\begin{equation}
	\fullket{z_{i_0}}_{|\Gamma} = e^{-i \Theta} \fullket{z_{i_0}}
\end{equation}
where $ \Theta $ is, in general, a matrix valued object which is a function of the labels of both the original puncture and the punctures enclosed by the loop:
\begin{equation}
	\Theta := \Theta(z_{i_0}; z_{i_1}, z_{i_2},\ldots,z_{i_k})
\end{equation}
Given an arbitrary two-dimensional lattice one would think that it should be possible to expand wave-functions on this lattice in terms of an overcomplete set of states spanned by all possible loops generated by moving one particle around all the other particles and repeating this procedure for all sites on the lattice. Conceivably it is possible to reverse this construction to allow us to determine a lattice structure for a set of points on a two-dimensional surface, given only the topological information encoded in their relative phases. We will assume that such a construction is indeed possible in the following.
\todo{In another \textbf{para or two} explain/argue for how information from holonomies can be translated into an effective lattice structure. \textbf{Hint:} Use the fact that \emph{given} a two-dimensional lattice, one can expand wave-functions on this lattice in a basis spanned by all possible loops generated by moving one particle around all the other particles \ldots.}


\subsection{Non-interacting many-body ground state}
\label{subsec:ground-state}

We would like to argue that the elementary quanta of geometry are tetrahedra which can be labeled by four points. Let us assume that a state of geometry represented by these points can be written as:
\begin{equation}
|\Psi>=|z_{1},z_{2};z_{3},z_{4}>\label{eqn:CondensateState}
\end{equation}
The above notation is meant to be reminiscent of the BCS formulation where pairs of electrons of equal and opposite momenta and spins (and near the fermi surface) scatter into another pair with equal and opposite
momenta and spins. This process can be represented as: $|k\uparrow,-k\downarrow>\rightarrow|k'\uparrow,-k'\downarrow>$.

The state $\Psi$ will have some symmetries under exchange of any two points. Under exchange the state should pick up a phase given by:
\begin{equation}
|z_{1}\dots z_{j}\dots z_{i}\dots z_{n}>=e^{-\imath\,\theta_{ij}}|z_{1}\dots z_{i}\dots z_{j}\dots z_{n}>\label{eqn:ExchangePhase}
\end{equation}
where we have changed our notation slightly so it applies to a state with an arbitrary number $n$ of punctures.

A physical state should be invariant under this exchange symmetry and should be analytic. A state of this form is then written as:
\begin{equation}
\Psi_{phys}=e^{+\imath\alpha\sum_{i<j}\theta_{ij}}\prod_{k<l}(z_{i}-z_{j})^{\alpha}
\label{eqn:GaugeInvariantState}
\end{equation}
\todo{{[}To do: Write out scalar-vector-tensor decomposition of metric perturbation around flat space and relate those terms to tetrads.{]}}

\subsection{Frodden-Perez-Ghosh + Interactions = Quantum Horizon}
\label{subsec:fpg-hamiltonian}

Given the phenomenological connections between the QHE and BHE and the proposed form of the horizon wavefunctions \eqref{eqn:GaugeInvariantState}, it would be satisfying if we could associate a notion of a \emph{hamiltonian} with the horizon degrees of freedom. In the context of general relativity this would imply the existence of a \emph{local} notion of energy associated with a black hole horizon. Until very recently the best one could do was assign a quasi-local energy function to the horizon using the generalized second law of black hole thermodynamics \textbf{references}: insert equation (1) from FPG \cite{Frodden2011A-local}.

Recent work done by Frodden, Ghosh and Perez \cite{Frodden2011A-local} and followup work by Bianchi \cite{Bianchi2012Entropy} and Bianchi and Wieland \cite{Bianchi2012Horizon} shows that it is natural to identify the area of the horizon with the local energy of the horizon. This development allows us to conjecture that one should be able to write down a hamiltonian for the horizon degrees of freedom in which the area operator $\vect{j_i}^{\dagger} \vect{j_i}$ at any puncture plays the role of the kinetic energy:
\begin{equation}\label{eqn:hamiltonian-kinetic}
	H_k^0 = - \sum_{i > j} j_{i,\alpha}^{\dagger} j_{j,\alpha} \delta^{ij} =  - \sum_{i} j^\dagger_{i,x} j_{i,x} + j^\dagger_{i,y} j_{i,y} + j^\dagger_{i,z} j_{i,z} 
\end{equation}
where the subscripts $i, \alpha$ on the ladder operators label the puncture and the component of the spin ($x,y,z$) at that puncture, respectively and the sum is over all pairs of punctures. The purpose of using the Kronecker delta notation in the second expression above is to suggest that \emph{in general} the Hamiltonian need not be isotropic or even diagonal and could have off-diagonal terms representing interactions between pairs of spins, as in:
\begin{equation}
			H_k = - \sum_{i > j} j_{i,\alpha}^{\dagger} j_{j,\beta} t^{ij;\alpha \beta}
\end{equation}
where $t^{ij;\alpha\beta}$, which need not be diagonal. The diagonal elements of $t_{ij;\alpha\beta}$ determine the kinetic energy at each site and the off-diagonal elements determine the strength of the coupling between different sites, thus it is referred to as the ``hopping matrix'' as it determines the amplitudes for a particle to ``hop'' between pairs of sites.

$H_k^0$ describes the non-interacting sector of the system of punctures on a surface. While in a macroscopic limit such a system will have an entropy of the Bekenstein-Hawking form, it does not include inter-particle interactions. One might speculate, that by considering Hamiltonians of the general type $H_k$ - which subsumes the Heisenberg, XX and other such models - one can model surfaces which have properties beyond those encountered in the behavior of horizons in general relativity.

\section{Conclusion}\label{sec:conclusion}

A quantum theory of horizons does not tell us how to make contact with the picture we started with where horizons are embedded in a $3+1$ dimensional bulk geometry. One might hope that in order to reconstruct something resembling $3+1$ geometry from the $2+1$ dimensional quantum theory, we should look towards holography and the gauge-gravity duality. It turns out that there does exist a compelling body of work \cite{Vidal2006Entanglement,Swingle2009Entanglement,Evenbly2011Tensor} based on a technique known as ``multi-scale entanglement renormalization'' or MERA for short which suggests that a macroscopic bulk geometry emerges when one considers correlations on larger and larger scales on the quantum horizon.

\subsection{``Quantum Distance'' and emergent geometry}\label{subsec:quantum-distance}

A readable review of such ideas can be found in \cite{Van-Raamsdonk2010Building} and a concrete proposal for how a higher-dimensional metric arises from a lower-dimensional quantum system is given in \cite{Jonckheere2011Geometry,Jonckheere2012Curvature}. The proposed methodology for determining a coarse-grained ``distance'' and hence a ``metric'' on a spin-network can be best illustrated for the case where the graphs are one-dimensional. Consider a chain of spins with local (nearest neighbor and higher order) interactions between neighboring spins. One can define the ``information transfer capacity'' between two lattice sites $i$ and $j$ as the probability $p_{i,j}$ with which one can transfer a quantum state at site $i$ to the site $j$. Then we can define a distance functional $d(i,j)$ between pairs of sites by taking the logarithm of this probability:
\begin{equation}\label{eqn:information-distance}
	d_{i,j} = - \log (p_{i,j})
\end{equation}
Such a functional between pairs of sites, where sites with $p_{i,j}=1$ - perfect transmission fidelity - are identified with the same ``point'', satisfies the requirements of a distance measure:
\vspace{-15pt}
\begin{subequations}
	\begin{align}
		d_{i,j} & \geq 0 \quad & \mathrm{(non-negativity)}\\
		d_{i,j} & = 0 \,\, \iff \,  i \equiv j \quad & \mathrm{(distinguishability)} \\
		d_{i,j} & = d_{j,i} \quad & \mathrm{(symmetry)} \\
		d_{i,j} + d_{j,k} & \geq d_{i,k} \quad & \mathrm{(triangle inequality)}
	\end{align}
\end{subequations}

The existence of a such an \emph{information based} distance measure between lattice sites implies that we can endow the lattice with a metric and other, higher, geometrical constructs. This approach towards identifying a suitable coarse-grained geometry with a quantum spin-network has the advantage that it is grounded in the information theoretic language. The resulting geometric invariants such as the Ricci scalar will then naturally encode the \emph{information-theoretic} or \emph{entropic} characteristics of the underlying quantum geometry. Any dynamical theory of geometry constructed from such geometric invariants will then have a natural interpretation as a \emph{thermodynamical} theory in line with the expectations from previous investigations including \cite{Jacobson1994Black},\cite{Padmanabhan2002Classical,Padmanabhan2003Gravity,Padmanabhan2012Emergent} and \cite{Verlinde2010On-the-Origin}.

\appendix

\section{$\sltwoc$ transformations}\label{app:lorentz-group}

The generators of the $ \mathfrak{sl}(2,\mathbb{C}) $ Lie algebra are:
\begin{equation}\label{eqn:sl2c_algebra}
\begin{array}{cccc}
	\sigma_x = \left( \begin{array}{cc} 0 & 1 \\ 1 & 0 \end{array} \right) &
	\sigma_y = \left( \begin{array}{cc} 0 & i \\ -i & 0 \end{array} \right) &
	\sigma_z = \left( \begin{array}{cc} 1 & 0 \\ 0 & -1 \end{array} \right) &
	\sigma_t = \left( \begin{array}{cc} 1 & 0 \\ 0 & 1 \end{array} \right)
\end{array}
\end{equation}
Any point on the celestial sphere with spacetime co-ordinates $\{T,X,Y,Z\}$ (where $X^2 + Y^2 + Z^2 = T^2$) is mapped onto a point $ z = x + iy $ on the complex plane $ \mathbb{C} $ under this stereographic projection. Lorentz transformations acting on this spacetime point can then be identified with $\sltwoc$ transformations acting on the corresponding point $z \in \mathbb{C}$.

Any 4-vector $\vec{U}$ in Minkowski spacetime can be expanded in terms of its components in a local basis $ \{ \mf{e}_i \} $(or \emph{tetrad}) as in:
\begin{equation}
	\vec{U} = U^i \mf{e}_i
\end{equation}
Alternatively one can contract the components of the vector $\vec{U}$ with the basis of the $\mathfrak{sl}(2,\mb{C})$ algebra to get an element of this Lie algebra, as in:
\begin{equation}
	\vec{U}^{AA'} = U^i \sigma_i^{AA'} \equiv \left(
	\begin{array}{cc} U^t + U^z & U^x + i U^y \\
					  U^x - iU^y & U^t - U^z 													    \end{array} \right)
\end{equation}
\ignore{
\section{The Modular Group: $\sltwoz$}\label{app:modular-group}

$\sltwoz$ is the group of fractional, linear transformations of points $z$ on the complex plane. Under the action of an element of $\sltwoz$ we get:
\begin{equation}
	z \rightarrow \frac{az + b}{cz + d}
\end{equation}

where $a,b,c,d \in \mathbb{Z}$ and $ad-bc=1$. The corresponding element is the matrix:
\begin{equation}
\left(	\begin{array}{cc}
			a & b \\
			c & d
		\end{array} \right)
\end{equation}
}

\printbibliography

\end{document}